\documentclass[msom,nonblindrev]{informs4} % current default for manuscript submission
\OneAndAHalfSpacedXI % current default line spacing
\usepackage[normalem]{ulem}
\usepackage{xcolor}
\usepackage{hyperref}
\usepackage{xurl}

\usepackage{algorithm}
\usepackage{algpseudocode}
\usepackage{amsmath}

\usepackage{natbib}
 \bibpunct[, ]{(}{)}{,}{a}{}{,}%
 \def\bibfont{\small}%
 \def\BIBand{and}%

\TheoremsNumberedThrough     % Preferred (Theorem 1, Lemma 1, Theorem 2)
\ECRepeatTheorems

\EquationsNumberedThrough    % Default: (1), (2), ...
\MANUSCRIPTNO{}

\begin{document}
%%%%%%%%%%%%%%%%

% Outcomment only when entries are known. Otherwise leave as is and
%   default values will be used.
%\setcounter{page}{1}
%\VOLUME{00}%
%\NO{0}%
%\MONTH{Xxxxx}% (month or a similar seasonal id)
%\YEAR{0000}% e.g., 2005
%\FIRSTPAGE{000}%
%\LASTPAGE{000}%
%\SHORTYEAR{00}% shortened year (two-digit)
%\ISSUE{0000} %
%\LONGFIRSTPAGE{0001} %
%\DOI{10.1287/xxxx.0000.0000}%

% Author's names for the running heads
% Sample depending on the number of authors;
% \RUNAUTHOR{Jones}
% \RUNAUTHOR{Jones and Wilson}
% \RUNAUTHOR{Jones, Miller, and Wilson}
% \RUNAUTHOR{Jones et al.} % for four or more authors
% Enter authors following the given pattern:
\RUNAUTHOR{Freeman et al.}

% Title or shortened title suitable for running heads. Sample:
% \RUNTITLE{Bundling Information Goods of Decreasing Value}
% Enter the (shortened) title:
\RUNTITLE{Linking Multi-Site Ad Data to Aid Counter-Trafficking}

% Full title. Sample:
% \TITLE{Bundling Information Goods of Decreasing Value}
% Enter the full title:
\TITLE{Linking Multi-Site Sex Ad Data at the Individual Level to Aid Counter-Trafficking Efforts}

% Block of authors and their affiliations starts here:
% NOTE: Authors with same affiliation, if the order of authors allows,
%   should be entered in ONE field, separated by a comma.
%   \EMAIL field can be repeated if more than one author
\ARTICLEAUTHORS{%
\AUTHOR{Nickolas K. Freeman, Gregory J. Bott,  Burcu B. Keskin, Jason M. Parton, James J. Cochran}
\AFF{Department of Information Systems, Statistics, and Management Science,\\ The University of Alabama, Tuscaloosa, AL 35478\\ \EMAIL{freem028@ua.edu}} %, \URL{}}
% Enter all authors
} % end of the block

\ABSTRACT{\textbf{Problem definition}: The Internet facilitates sex trafficking through adult service websites (ASWs) that host online advertisements for sexual services (sex ads). Since the closure of the popular site Backpage.com, the ecosystem of ASWs has expanded to include multiple competing sites that are hosted outside US jurisdiction. Gaining intelligence for counter-trafficking efforts requires collecting, linking, and cleaning the data from multiple sites. However, high ad volumes, disparate data types, and the existence of generic and misappropriated data make this process challenging. \textbf{Academic/practical relevance}: We present an end-to-end process for linking sex ad data and filtering potentially erroneous links. Outputs of the developed process have been used to inform counter-trafficking operations that have helped identify more than 60 potential victims of sex trafficking, some of whom are getting help to transition out of \textit{the life}. \textbf{Methodology}: Our process leverages concepts and techniques from network science, information systems, and artificial intelligence to link ads across sites at the level of an individual or unique posting entity. Our approach is computationally efficient, allowing millions of ads to be processed in under an hour. \textbf{Results}: A key component of our process is an edge filtering procedure that identifies and removes potentially erroneous links in a graph representation of sex ad data. A comparison of the proposed process to an existing approach shows that our process is typically more computationally efficient and yields substantial increases in the number of individuals for which we can derive actionable intelligence. \textbf{Managerial implications}: The proposed process is an efficient and effective approach for transforming the high volumes of disparate data from sex ads into intelligence that can save lives. It has been refined over years of collaboration with practitioners and represents a strong foundation upon which further counter-trafficking tools can be built. 
}

% Sample
%\KEYWORDS{deterministic inventory theory; infinite linear programming duality;
%  existence of optimal policies; semi-Markov decision process; cyclic schedule}

% Fill in data. If unknown, outcomment the field
\KEYWORDS{sex trafficking; network science; anomaly detection; artificial intelligence}

\maketitle
\section{Introduction}\label{sec:introduction}

Sex trafficking refers to the use of force, fraud, or coercion to compel an individual to perform commercial sex services. The Internet has and continues to facilitate sex trafficking in the US via adult service websites (ASWs) that host commercial sexual advertisements (sex ads). Recent news references highlight the ongoing role of the Internet in facilitating sex trafficking via ASWs:
\begin{itemize}
    \item \cite{USAOMDNC} describes the sentencing of a Durham man to six life terms in prison for sex trafficking. The trafficker targeted women who were homeless or suffering from substance abuse problems. Five victims testified that they were photographed and advertised on two ASWs. They also testified that the sentenced individual and a co-defendant arranged commercial sex encounters and that the trafficker used violence and coercion to ``maintain order.''
    \item \cite{fox6milwaukee} describes the sentencing of a Milwaukee man for sex trafficking of two minors in Chicago and Wisconsin. The trafficker advertised the victims on two ASWs and arranged encounters to occur in hotel rooms.
    \item \cite{example3} describes a man pleading guilty to charges of sex trafficking of a minor. The trafficker posted ads for the victim on several ASWs, interacted with potential \textit{clients}, and scheduled encounters between the trafficked individual and clients.
\end{itemize}

In recent years, government agencies have engaged in a game of whack-a-mole, shuttering ASWs linked to sex trafficking activities only to have new sites quickly emerge to fill the resulting void. \cite{zeng2022internet} discuss the futility of these efforts, finding no causal impact that is attributable to the shutdown of two prominent sites in 2018. Instead, they found that users quickly moved to other sites hosted in locations outside of US jurisdiction. The persistence of the online commercial sex ecosystem, demonstrated by anecdotal evidence and empirical analyses, suggests it is here to stay. Consequently, organizations dedicated to combating sex trafficking will continue to need methods that transform the data available on these platforms into actionable intelligence. 

Not all ads posted on ASWs are associated with victims of sex trafficking, and distinguishing advertisements featuring trafficking victims from those posted voluntarily or created as scams is a significant challenge. A large body of research is devoted to detecting sex trafficking victims based solely on the attributes of sex ads, i.e., keywords, emojis, and image attributes. However, a critical issue is the absence of large-scale, verifiable ground truth data. Even when law enforcement organizations can identify and isolate victims, establishing definitive evidence of trafficking requires significant time and resources, and victims often deny their victimization due to trauma or fear.

In addition to the challenge of identifying trafficking victims, the process of transforming sex ad data into actionable insights requires addressing several technical obstacles. First, sex ads use various types of data, including text, images, and numbers. Thus, data transformation pipelines must handle and link this multimodal data. Second, although ASWs typically convey the same core information, the ASW ecosystem includes many sites with different formats that provide data with slight variations from others. Third, the high volume of ads created daily exceeds the ability of most law enforcement organizations (LEOs) and non-profit organizations (NPOs) to harvest, store, and transform this data into meaningful data products. Fourth, since providing and consuming commercial sex are illegal in most of the continental US, scammers target ASWs extensively, adding significant noise to the data. Finally, even legitimate ads may contribute noise in the form of erroneous connections that arise from the use of generic or misappropriated data.

Since 2019, our research has focused on developing a systematic process for collecting and analyzing sex ads to support anti-trafficking efforts. Our \textit{ad-linking} process ($i$) uses a graph representation to link the diverse data included in ads, ($ii$) filters a \textit{giant component} that emerges during graph construction and includes erroneous links between a large proportion of the data; and ($iii$) and transforms the filtered graph representation into actionable data products to inform the efforts of LEO and NPO partners. These data products provide quantitative analyses of advertisement frequency within specific geographic regions, such as cities or counties, along with associated ad URLs and phone numbers. Images are available on request, which may be blurred to avoid traumatizing NPO volunteers, some of whom are survivors of sexual exploitation. These data products have supported LEO and NPO operations, resulting in the identification of more than 60 potential sex trafficking victims. We refer to these individuals as ``potential victims'' because convictions are difficult due to the common requirement of victim testimony. Given the emotional, physical, and psychological trauma associated with exploitation, it is often difficult to secure such testimony, even if trafficking has occurred.

Our ad-linking process builds on that described in \cite{freeman2022collaborating}, addressing several limitations and incorporating significant enhancements. Specifically, our process ($i$) incorporates recent advances in language modeling for AWS data, ($ii$) leverages connected components from a graph representation of the data in a novel fashion to generate training examples for a \textit{same user} classifier, and ($iii$) implements a novel approach for filtering erroneous data links without data loss. Moreover, we disclose as much detail as possible regarding our methods and computational setup to aid reproducibility. Using a large open ASW dataset, we show that the proposed approach typically offers substantial improvements in terms of output quality and computational time.

The remainder of the paper is structured as follows. Section \ref{sec:related_work} describes related research on linking data from ASWs. Section \ref{sec:background} summarizes key concepts and techniques from graph theory/network science, image processing, and machine learning/artificial intelligence that we incorporate in our methodology. Section \ref{sec:ad_linking_process} provides a detailed description of the various components of our ad-linking process. Section \ref{sec:gc_edge_removal} discusses our approach to detecting and removing erroneous links in the giant component that arises in the graph representation of the data and presents experimental results demonstrating its efficacy. Section \ref{sec:conclusion} concludes the work, acknowledges limitations, and suggests directions for future research.
\section{Related Work}\label{sec:related_work}

This paper presents an end-to-end process to help identify victims of sex trafficking by linking sex ad data at the level of an individual or unique posting entity (e.g., escort agency, massage parlor, pimp). A significant component of our approach is a procedure for filtering the giant component that arises in graph representations of the data due to generic posts, generic images, and misappropriated images in these datasets. In this section, we provide an overview of related works that use analytical methods to link and derive insights from sex ad data. We also discuss some related research on filtering giant components in graphs. We conclude the section by discussing specific contributions of this research.

\subsection{Analytical Methods for Deriving Insights from Sex Ads}
\cite{li2018detection} presents one of the earliest studies on linking sex ad data. The authors propose an unsupervised template matching algorithm that detects \textit{organizations} operating on ASWs, where the term \textit{organization} refers to ``a single individual or a network of individuals who are trafficking a specific set of victims.'' Analogously, our ad-linking process extracts connected components of a graph representation of ad data corresponding to an individual or unique posting entity. Their matching algorithm generates embeddings for sex ad texts, clusters the embeddings using Hierarchical Density-Based Spatial Clustering of Applications with Noise (HDBSCAN) \citep{mcinnes2017hdbscan}, and extracts word groups or \textit{templates} that exhibit a high degree of collocation in the clustered documents. Applying their algorithm to a dataset of nearly 40 million ads collected from the (now defunct) site \textit{Backpage.com} demonstrates it generates ($i$) linkages similar to those deduced from a graph representation based on posted phone numbers and ($ii$) additional linkages that cannot be constructed using only phone numbers. This research represents an important first step in showing how sex ad features can be leveraged to detect similarity. 

\cite{zhu2019identification} use machine learning to detect phrases indicative of sex trafficking in sex ad data. Specifically, they apply support vector machines on normalized ad text from the Trafficking10k dataset (see \cite{tong2017combating} for more information) to classify ads into various risk categories. \cite{lee2021infoshield} describes INFOSHIELD, which detects patterns in sex ad text. The procedure uses term frequency-inverse document frequency (TF-IDF) to extract distinguishing keywords from posts and then constructs a bipartite graph that captures document-phrase relationships. The connected components of this graph represent a \textit{coarse} clustering, which is refined in a \textit{fine} clustering phase. The authors test the approach using open Twitter bot data (see \cite{cresci2017paradigm} for more details) and the Trafficking10k dataset.

Similar to the previously described research, we use a graph representation for the sex ad data. However, we use a transformer-based neural network model to capture and compare the semantic meanings of posts instead of TF-IDF. We also use perceptual hashes (pHashes) to incorporate images into our ad-linking procedure. \cite{keskin2021cracking} are the first researchers to demonstrate the use of perceptual hashing techniques for linking sex ads using text and images. They extract spatiotemporal patterns from linked ad groups and use them to simulate future movements. \cite{freeman2022collaborating} extends the work of \cite{keskin2021cracking} by describing an end-to-end process for collecting and grouping sex data. They also use a graph representation of sex ad data and observe that generic text, generic images, and misappropriated images can create erroneous links, leading to the formation of large connected components in the graph. They propose a method for filtering these large components based on betweenness centrality, but note that sampling techniques are required to overcome computational challenges. Although the authors are correct that erroneous links can form in the graph representation, they fail to recognize that this leads to a well-studied phenomenon in network science: the emergence of a giant component.

\subsection{Filtering the Giant Component in Graphs}
A significant issue we encounter when constructing the graph representation of ASW data is the presence of a giant component. In network science, a giant component is a connected component of a graph whose size scales proportional to the number of vertices in the graph \citep{sood2023existence}. Although the existence of a giant component is often considered a useful feature of a network, there are cases where decomposing it offers important insights into the network. \cite{dianati2016} discusses techniques and applications for such decomposition methods, which the authors refer to as \textit{graph pruning} methods. One use case, which also applies to illicit activities, is presented in \cite{Husain2020}. Specifically, the authors use a graph representation to study spatiotemporal patterns in data related to terrorism. They use the Global Terrorism Database, which contains news articles on terrorism events from around the world, to construct a graph connecting terrorists and their targets. They find that the constructed graph representation includes a giant component that links approximately 92\% of all vertices. They use a disparity filtering method to filter edges and extract the \textit{backbone} of the graph.

\subsection{Contribution}
The closest research to ours is \cite{freeman2022collaborating}. Although their goal is similar, there are several limitations with their proposed pipeline. First, the authors fail to recognize the emergence of a giant component in the graph representation of ASW data, which leads them to devise filtering methods that apply to other components unnecessarily. Second, the procedure proposed for identifying and filtering potentially erroneous links in \cite{freeman2022collaborating} ($i$) does not leverage the data associated with vertices, ($ii$) can lead to data loss, and ($iii$) requires approximation to overcome computational challenges. Specifically, the authors use estimates of betweenness centrality (BC) to identify and remove vertices that exceed a selected percentile threshold in the graph representation. BC computation is very expensive. Thus, the authors are required to approximate the BC values for all vertices in a particular component based on shortest-path computations among a subset of nodes. Regarding data loss, filtering graph vertices based on extreme BC percentile should result in the removal of only a few vertices with high connectivity. However, as we will show in our experimentation, different site structures or approximation issues can lead to much more substantial data loss. A final limitation of \cite{freeman2022collaborating} is that it lacks several details that would be necessary for replication.

Our ad-linking process makes several contributions and addresses limitations of the process presented in \cite{freeman2022collaborating}. First, we identify the emergence of a giant component, connecting the challenges of working with sex ad data to a broader phenomenon in the network science literature. Second, our approach to filtering potentially erroneous edges in the giant component is data-driven, leveraging recent advances in language modeling for AWS data to identify and remove such connections based on classifying the similarity of ad text pairs. Unlike prior methods, the present filtering procedure does not result in data loss. A major challenge of training such a classifier is the lack of ground truth. We present a novel approach to generate reliable training data for the task in a dynamic fashion that is robust to changes in the underlying data used to construct the graph representation. This approach is not specific to our application and can be applied in other contexts where filtering the giant component provides insights into its underlying structure. Finally, this research takes major steps towards reproducibility, including detailed descriptions for all steps and leveraging a large and open ASW dataset \cite{Freeman-2025-multisite} in experiments that demonstrate its efficacy in terms of output quality and computational time. Ultimately, the outlined process opens the door for new research directions by providing a thorough, rigorous, and well-documented procedure for extracting data corresponding to posting entities represented in ASW data. 
\section{Background}\label{sec:background}
This section briefly describes concepts and techniques we employ from graph theory/network science, image processing, and machine learning/artificial intelligence. Our intent is not to provide an overarching review of any area, concept, or technique, but to highlight aspects that will aid in understanding our description of the developed ad-linking process.

\subsection{Graph Theory and Network Science Fundamentals}\label{subsec:graph_fundamentals}
Network science refers to the application of mathematical graph theory to problems across various disciplines, where graph theory focuses on mathematical approaches for modeling relationships between objects \citep{lewis2011network}. A graph $G(V, E)$ is defined by a set of vertices $V$ that are connected by edges that form a set $E$. These edges can be directed or undirected, depending on the context. For example, vertices in a family tree graph may represent individuals, and directed edges can indicate who is a descendant of whom. Edges can also have weights that represent meaningful attributes of the relationship between the two connected vertices, such as distance or capacity. A path, $P_{uv}$, refers to an ordered collection of vertices connecting an origin, or source, vertex $u \in V$ to a destination, or target, vertex $v \in V$. The path's length is the number of edges the path includes. A vertex is said to be \textit{reachable} from another if a path connecting the two vertices exists \citep{borner2007network}.

Our ad-linking process represents sex ad data as a graph and extracts groups of vertices that are reachable from one another. The groups are called components in graph theory and network science terminology. A common phenomenon in the study of real-world graphs is the emergence of a giant component, which is a component whose size scales with the number of vertices in the graph \citep{janson1993birth}. When a giant component exists, a significant proportion of the graph vertices will belong to this component \citep{borner2007network}. In our application, the presence of generic and misappropriated data leads to the emergence of the giant component. Our ad-linking process includes an edge filtering procedure that specifically seeks to identify and filter such erroneous edges in the giant component. Filtering the giant component allows us to decompose it into additional components, each likely to correspond to an individual or unique posting entity for which we can derive actionable intelligence from. 

\subsection{Hashing Techniques - Cryptographic and Perceptual}
Sex ads include text, contact information (often represented as strings), and images. Individuals or posting entities tend to reuse each of these data elements. Thus, we need efficient ways to represent and match them across ads and ASWs. Achieving this is most complicated for images.

In computer science, a hashing function maps data records to a lower-dimensional space such that two pieces of identical data generate the same hash value. Ideally, these functions create simple encodings that can verify more complex data or programs. Various hashing functions have been developed for different applications \citep{chi2017hashing}, with cybersecurity being one of the most well-known applications. 

Cryptographic hashing refers to security-oriented methods designed to be one-way and difficult or infeasible to reverse. They are also designed to ensure that small content changes result in different hash values. A common application of cryptographic hashes is verifying downloaded software. Software providers compute and publish hash values using specific algorithms, such as Message Digest (e.g., MD5) or Secure Hash Algorithm (e.g., SHA-256). Users who download the software can apply the same algorithm to verify that their hash matches the one published, with different values potentially indicating corrupt software.

Perceptual hashing techniques have a similar goal of representing complex data in a simpler format, but are designed to work with images. Specifically, digital images are represented as arrays of pixel values, and perceptual hashing techniques determine whether two images are fundamentally the same based on their content, rather than making strict pixel-to-pixel comparisons. This is crucial because slight image variations can arise from cropping, applying filters, compression, or other modifications. Perceptual hashing techniques are commonly used to help identify instances of copyright infringement or online child sexual abuse materials (CSAM). We refer readers interested in learning more about perceptual hashing techniques and their applications to \cite{farid2021overview}.

\subsection{Transformer Models and Sentence Embeddings}
As noted earlier, our graph representation for sex ad data is prone to the emergence of a giant component that arises from generic and misappropriated data. Our ad-linking process includes a procedure to identify and remove erroneous edges within the giant component. This filtering procedure leverages transformer neural network models tailored for ASW applications for these comparisons.

The transformer neural network architecture, introduced by \cite{vaswani2017attention}, utilizes an attention mechanism that allows the model to focus on important contextual elements during training. Before transformers, Recurrent Neural Networks (RNNs) were the dominant architecture for text processing. RNNs process text sequences serially, making them slow and computationally challenging for large context windows. The transformer architecture and its attention mechanism have removed significant barriers, enabling rapid advances in neural network applications for natural language and other areas \citep{incitti2023beyond}. 

One area where transformers have been successfully applied is developing sentence embedding and classification models. These models transform sentences into numerical vector representations that can be compared to identify sentences with similar contexts. The open-source community has emerged as a primary host of pre-trained transformer models. Platforms like \href{https://huggingface.co}{Hugging Face} host repositories of models and publish libraries that enable easy Python integration. On Hugging Face, contributors ranging from large technology companies to individuals share models and documentation that allow users to use pre-trained models directly or \textit{fine-tune} them for specific applications.

Several key concepts are important when applying transformer models to natural language problems:
\begin{hitemize}
\item[ - \textbf{Foundation Models}:] The term \textit{foundation model} refers to a model trained on broad, diverse datasets to provide a \textit{foundation} for more specialized models \cite{bommasani2022opportunitiesrisksfoundationmodels}. Some well-known foundation model families are \texttt{BERT} \citep{devlin2019bert}, \texttt{RoBERTa} \citep{liu2019roberta}, \texttt{GPT} \citep{radford2018improving}, and \texttt{LLaMa} \citep{touvron2023llama}. Foundation models learn vector representations for \textit{tokens}, which are basic units of text that can correspond to words, subwords, characters, or bytes. Different models employ different tokenization approaches. For example, \texttt{BERT} uses \textit{WordPiece} tokenization, while \texttt{GPT} uses \textit{Byte-Pair Encoding} tokenization. It is important to note that the vocabulary of tokens that a particular language model utilizes is fixed. If a foundation model encounters a token that is not in its vocabulary, it is mapped to special tokens that represent it as unknown, e.g., the \texttt{[UNK]} token in BERT. Understanding how unknown tokens are handled is crucial when considering ASW data, which uses emojis and slang that may not be part of the general-purpose vocabulary of mainstream foundation models.

\item[ - \textbf{Transformer Components}:] The transformer consists of encoder and a decoder components \citep{vaswani2017attention}. The encoder learns how to represent input text as vector embeddings that capture context from other tokens in an input sequence. The decoder learns to generate tokens autoregressively based on preceding tokens. Foundation models differ in the transformer components used. For example, \texttt{BERT} is an \textit{encoder-only} model that can generate contextualized token embeddings based on all other tokens in an input sequence \citep{qorib-etal-2024-decoder}. \texttt{BERT} variants are commonly fine-tuned for tasks that require understanding an entire text sequence, such as text classification and named entity recognition. \texttt{GPT} variants are \textit{decoder-only} models that are typically fine-tuned for tasks that rely on text generation. \texttt{T5} utilizes both the encoder and decoder components, and is commonly fine-tuned for tasks where text understanding and generation are important, such as question answering and text summarization.

\item[ - \textbf{Fine-Tuning}:] Foundation models are designed to exhibit broad text understanding for assessing text similarity or generating text. Using foundation models for a specific task, such as classification or named entity recognition, requires fine-tuning on a labeled dataset \cite{ding2023parameter}. Some tasks require an additional step of training a model that learns to aggregate token representations for a text sequence into a single embedding that can then be used to generate features for a classifier or clustering algorithm. An example of such a model is a sentence transformer model \citep{reimers2019sentence}.
\end{hitemize}

Our research utilizes a sentence transformer model specifically fine-tuned for computing similarities between post texts found in sex ads. \cite{freeman2025languagemodel} describes the training process, which includes custom pre-training of a \texttt{BERT} variant on a dataset of more than 19 million ASW posts and fine-tuning a sentence transformer on a dataset of 4 million ASW post triplets using a \textit{contrastive learning} task. We use this fine-tuned sentence transformer to generate numerical vector representations for post texts. We compute the similarity between post text pairs using cosine similarity, a standard method for comparing sentence embeddings \citep{reimers2019sentence}. We employ a simple binary classification model to translate the similarity scores into predictions about whether the same individual created both posts. 
\section{Ad-Linking Process}\label{sec:ad_linking_process}
\begin{figure}[htbp]
\FIGURE
{\includegraphics[width=\textwidth]{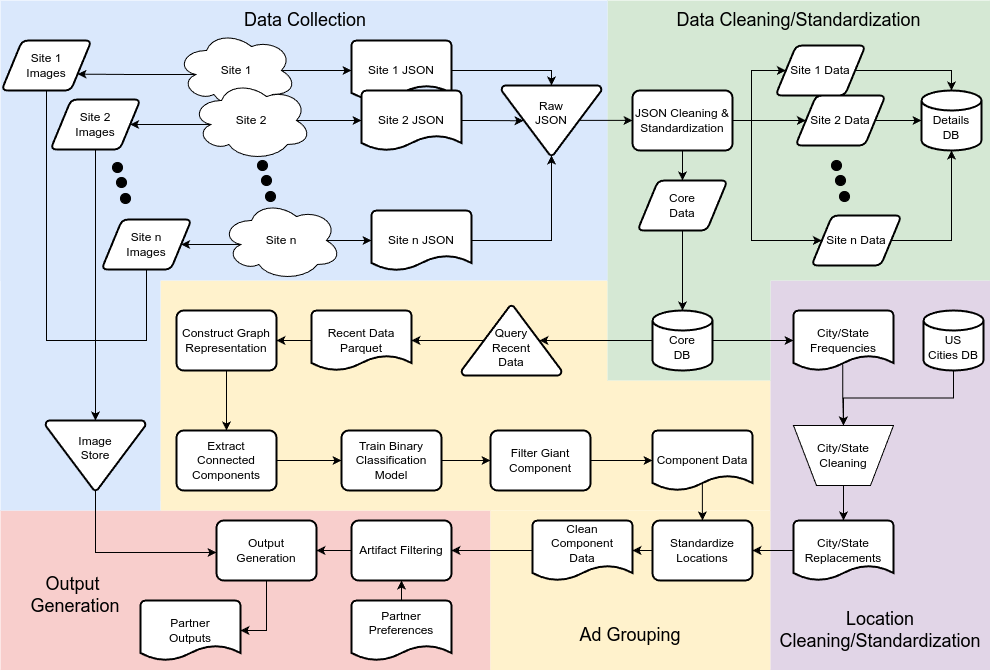}}
{Ad-Linking Pipeline\label{fig:ad_grouping_pipeline}}
{}
\end{figure}

This section describes our process for linking ads at the level of an individual or unique posting entity. The process performs five distinct tasks: ($i$) data collection, ($ii$) data cleaning and standardization, ($iii$) location cleaning and standardization, ($iv$) ad grouping, and $v$) output generation. Figure \ref{fig:ad_grouping_pipeline} provides a flowchart depicting the complete process. Each task is described in detail in the following subsections.

\subsection{Data Collection}\label{subsec:data_collection}
We use custom web scraping applications to collect data from ASWs. Because the ecosystem of ASWs is constantly evolving and the structure and format of each site differ, we develop specific data collection applications for each one. We add new applications for emerging sites, discontinue applications for inactive sites, and modify existing applications as sites change. 

Each application collects all text and image information for newly posted ads. We use unique identifiers, such as ad IDs or unique combinations of extracted features (like text and post date), to ensure ads are only collected once. We also utilize caching to ensure that we only collect each unique image once. All text information is stored in a JSON format, and we use cryptographic and perceptual hashing techniques to represent associated images in the JSON data. Copies of each unique image encountered are downloaded to a dedicated image store. 

\subsection{Data Cleaning and Standardization}\label{subsec:data_cleaning_and_standardization}
The data cleaning and standardization task serves two purposes: ($i$) to extract a consistent subset of information from all sites for ad linking and ($ii$) to map all data to a set of standardized labels, allowing comparisons between sites. We refer to the data extracted in ($i$) as \textit{core} data, which is stored in a relational database (Core DB). The data mapped in ($ii$) is referred to as \textit{detail} data and stored in a separate relational database (Details DB). 

The core data includes the following data fields for each ad, which are collected from all sites:
\begin{hitemize}
    \item Unique ID - unique identifier for ad,
    \item Post - heading text displayed with ad,
    \item Site - the ASW from which the ad was collected,
    \item Post Date - the date the ad was posted or first observed,
    \item Target City - the city the ad targeted,
    \item Target State - the state the ad targeted,
    \item Location Details - any freeform text providing additional information regarding the location,
    \item Phash-16 Values - a list of 16-bit pHash values for images posted with the ad and
    \item Contact Information - contact information provided with the ad.
\end{hitemize}
The detail data varies significantly among sites and includes additional information about the provider, such as hair and eye color, rates, touring schedules, and even client reviews.

\subsection{Location Cleaning and Standardization}\label{subsec:location_cleaning_and_standadrization}
The next phase of our ad-linking process focuses on cleaning and standardizing cities and states represented in the core data. Most ASWs allow posters to select from a limited set of cities and states when specifying the target location for their ad. This creates two significant challenges. First, the set of locations varies among sites. Second, the available locations may differ in their granularity and spelling of location names. Examples of these issues are one site allowing ad creators to select ``Dallas, Texas'' or ``Fort Worth, Texas'' whereas another only allows ``Dallas/Fort Worth, Texas.'' Moreover, ``Fort Worth'' can be represented as ``Fort Worth,'' ``Ft Worth,'' or ``Ft. Worth,'' each interpreted as a different string by a computer.

To address these issues, our process maps the city/state pairs associated with ads to locations specified in Version 1.6 of the \textit{Pro} US cities database made available by \textit{simplemaps} \citep{simplemaps}. We begin by first querying the Core DB to retrieve a dataset showing the number of URLs collected for each unique city-state pair. We then combine the city and state strings and attempt to match the combined string to an entry in the US cities database. In cases where we successfully match the city-state string to an entry in the US cities database, we join the county name and Federal Information Processing Standards (FIPS) code from the US cities database to the ad data. FIPS codes are standardized numeric or alphanumeric codes issued by the US government to uniquely identify geographic entities, such as states, counties, cities, and other administrative divisions. For cases where no match is made, we sort the data by the number of affected URLs and manually specify appropriate matches, prioritizing locations that will affect the most URLs. 

\begin{figure}[htbp]
\FIGURE
{\includegraphics[width=\textwidth]{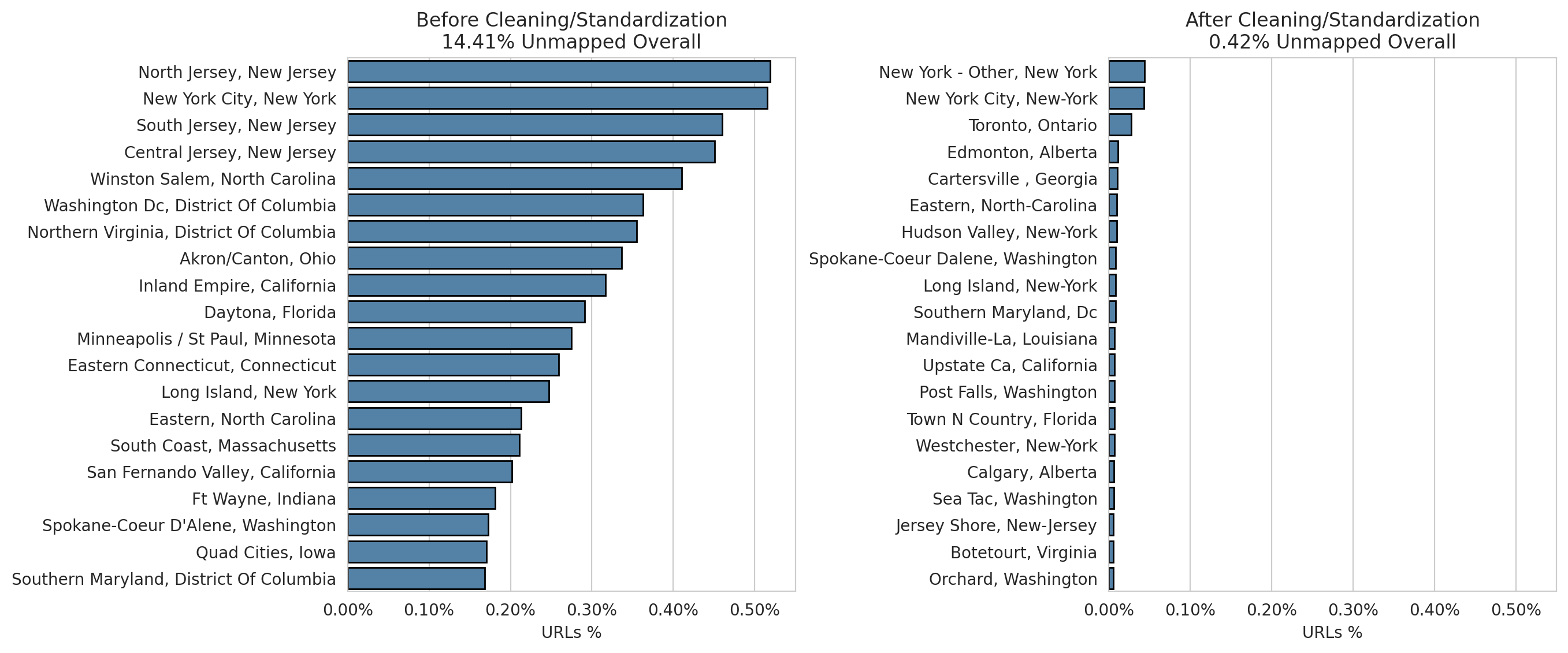}}
{County Mapping Examples\label{fig:county_mapping_examples}}
{}
\end{figure}

Our matching process is imperfect because we cannot specify appropriate matches for all cases. However, our efforts reduce the number of URLs affected by unmatched city/state pairs from approximately 15\% to less than 1\%. Figure \ref{fig:county_mapping_examples} provides additional details on the impacts of our manual location cleaning and standardization process. The left subplot shows the top 20 unmatched locations as a percentage of affected URLs before cleaning and standardization, while the right subplot shows the top 20 unmatched locations after cleaning and standardization. The output of the location cleaning and standardization process is a dataset that maps the observed city/state strings to cleaned representations with county details. 

\subsection{Ad Grouping}\label{subsec:ad_grouping}
Ad grouping begins by querying recent core data, which is saved to disk as a Parquet file. Parquet is an open-source file format designed to handle flat, columnar data. It is known for efficient data compression and its ability to handle various encoding types. Typically, we work with core data corresponding to ads posted or observed within 90 days of execution. This timeframe is based on feedback from our partners and can be adjusted as needed. 

After querying, we construct an unweighted and undirected graph representation of the data where vertices correspond to unique artifacts, such as post text or pHash values. Graph edges identify links between artifacts that appear together in an ad. A typical ad consists of a single post, multiple images, and can contain no contact information (some sites include a ``Message Me'' capability), one piece of contact information (e.g., a phone number), or multiple pieces of contact information. When constructing the graph representation, we represent each ad using the post text value as the central vertex connected to various pHash and contact information vertices. When multiple ads use the same artifact, links among ads create a larger graph structure. 

\begin{figure}[htbp]
\FIGURE
{\includegraphics[width=0.8\textwidth]{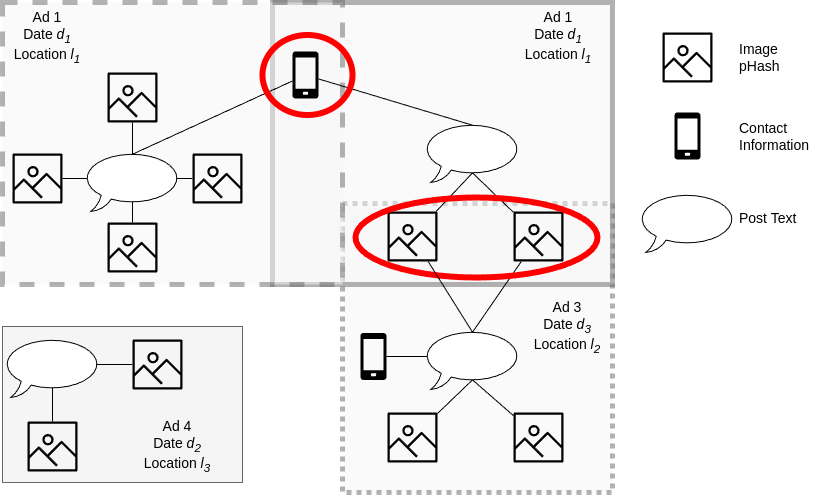}}
{Graph Representation Example\label{fig:graph_example}}
{The \textit{Post Text} vertices correspond to the raw post text harvested from an ad.}
\end{figure}

Figure \ref{fig:graph_example} shows a graph representation for a simple example with four ads, where ads 1, 2, and 3 are linked because they share common artifacts. Specifically, ads 1 and 2 share a common contact information vertex, e.g., the two ads used the same phone number. Also, ads 2 and 3 share two images. Ellipses are used to identify the shared artifacts. Due to the described overlaps, the artifacts associated with ads 1, 2, and 3 form a single connected component. Ad 4 has no artifacts that overlap with other ads; thus, the artifacts associated with ad 4 form another connected component. Note that we also know the post date and target location for all ads. When an individual posts the same ad multiple times, i.e., with identical post text, contact information, and images, we also have records of all post dates and target locations in the data extracted from the core DB.

Once the graph is constructed, we extract all connected components, each corresponding to a set of artifacts that can be linked via their usage across all ads. A significant issue we encounter at this point is the emergence of a giant component, which can largely be attributed to: ($i$) generic post texts and images (e.g., ``Available Now'' or the OnlyFans logo) and ($ii$) misappropriated images where a provider creates an ad using an image found elsewhere. The giant component poses a significant problem as it links artifacts associated with many providers. Without filtering the edges in the giant component and eliminating erroneous connections, the data it links is too noisy for actionable insights and must be disregarded.

\begin{figure}[htbp]
\FIGURE
{\includegraphics[width=\textwidth]{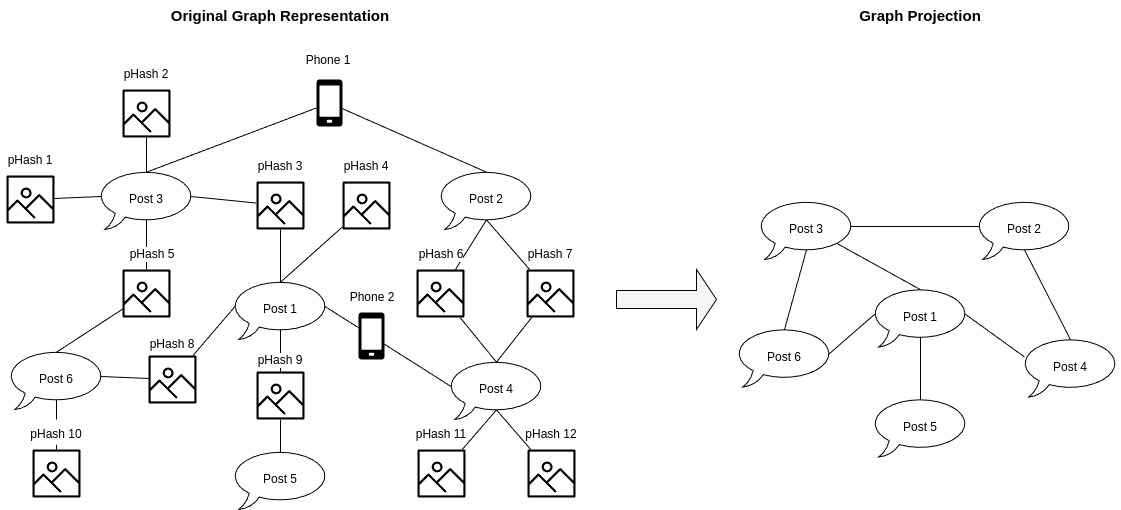}}
{Giant Component Filtering - Graph Projection\label{fig:GC_decomposition1}}
{}
\end{figure}

Our ad-linking process includes a giant component filtering procedure that leverages a prediction model to detect and remove likely erroneous edges from the giant component. Section \ref{sec:gc_edge_removal} provides a detailed explanation of the procedure. For now, we provide a high-level overview of how the procedure breaks down the giant component into smaller components likely to be associated with an individual or a unique posting entity. The first step is to extract the giant component from the graph representation of the data and project it so that it only includes vertices associated with posts. Edges in the projection exist between pairs of vertices corresponding to post text values that were connected by \textit{pHash} or \textit{contact information} vertices in the original graph representation. Figure \ref{fig:GC_decomposition1} depicts this graph projection step, where images correspond to pHash values and phones correspond to contact information values.

\begin{figure}[htbp]
\FIGURE
{\includegraphics[width=0.8\textwidth]{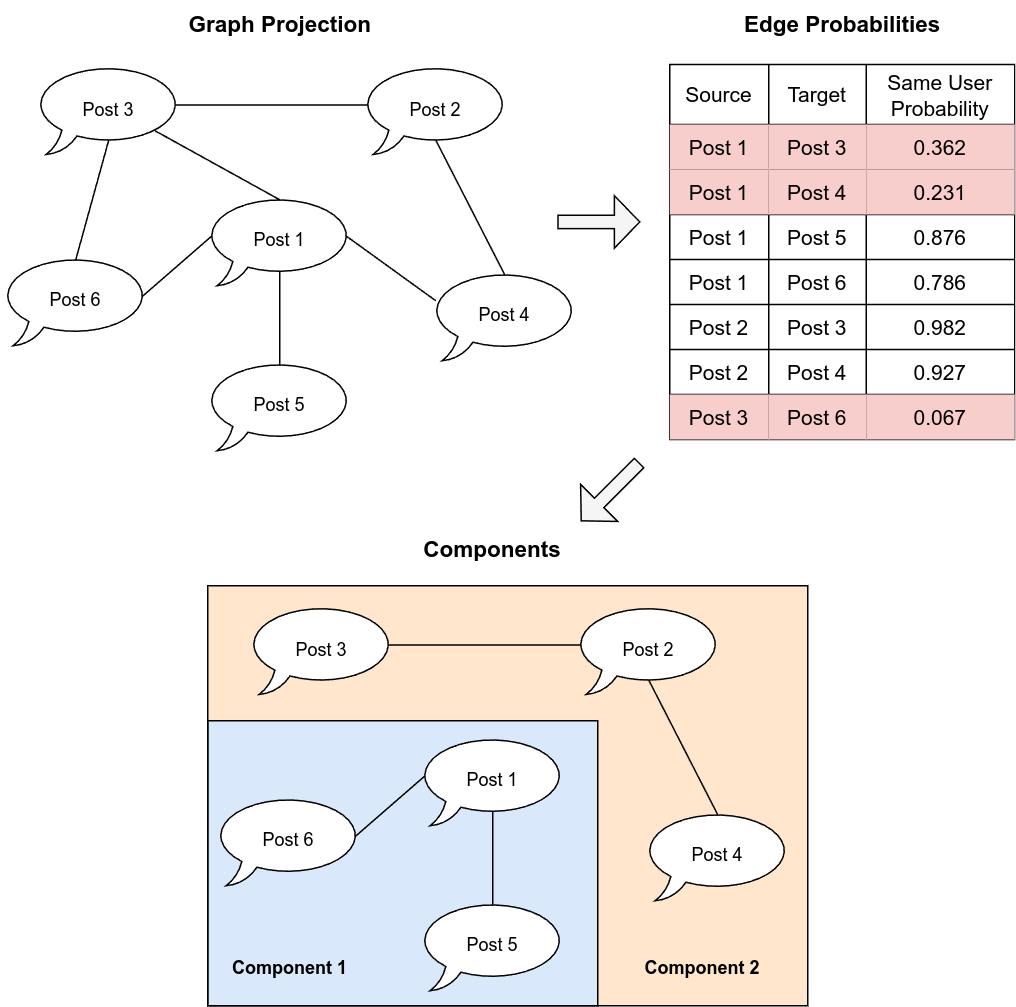}}
{Giant Component Filtering - Edge Removal\label{fig:GC_decomposition2}}
{}
\end{figure}

Next, we iterate over all edges in the graph projection and use a \textit{binary classification model} to assess the probability that the two posts connected by an edge were created by the same user or posting entity. We refer to this probability as the \textit{same user probability}. A user-specified threshold is used to identify edges to remove from the graph projection. Specifically, if the same user probability value for two linked posts is less than the threshold, the edge is flagged for removal. Flagged edges are removed after iterating over all edges in the projection, leaving components corresponding to connected posts. Figure \ref{fig:GC_decomposition2} illustrates how this procedure filters the graph projection from Figure \ref{fig:GC_decomposition1} into two distinct components of connected posts when a threshold of 0.5 is applied.

Components extracted from the filtered projection of the giant component allow us to map each post text value associated with a vertex in the giant component to a unique component identifier. Additionally, each component not included in the giant component can also be mapped to unique component identifiers. These unique identifiers are mapped back to the original data, along with county names and FIPS codes from the location cleaning/standardization phase. This data is written to a \textit{clean component data} file for use in the next phase. A key thing to note about the described filtering procedure is that it results in no data loss. Instead, we are simply removing associations between post text values.

\subsection{Output Generation}\label{subsec:output_generation}
\begin{figure}[htbp]
\FIGURE
{\includegraphics[width=\textwidth]{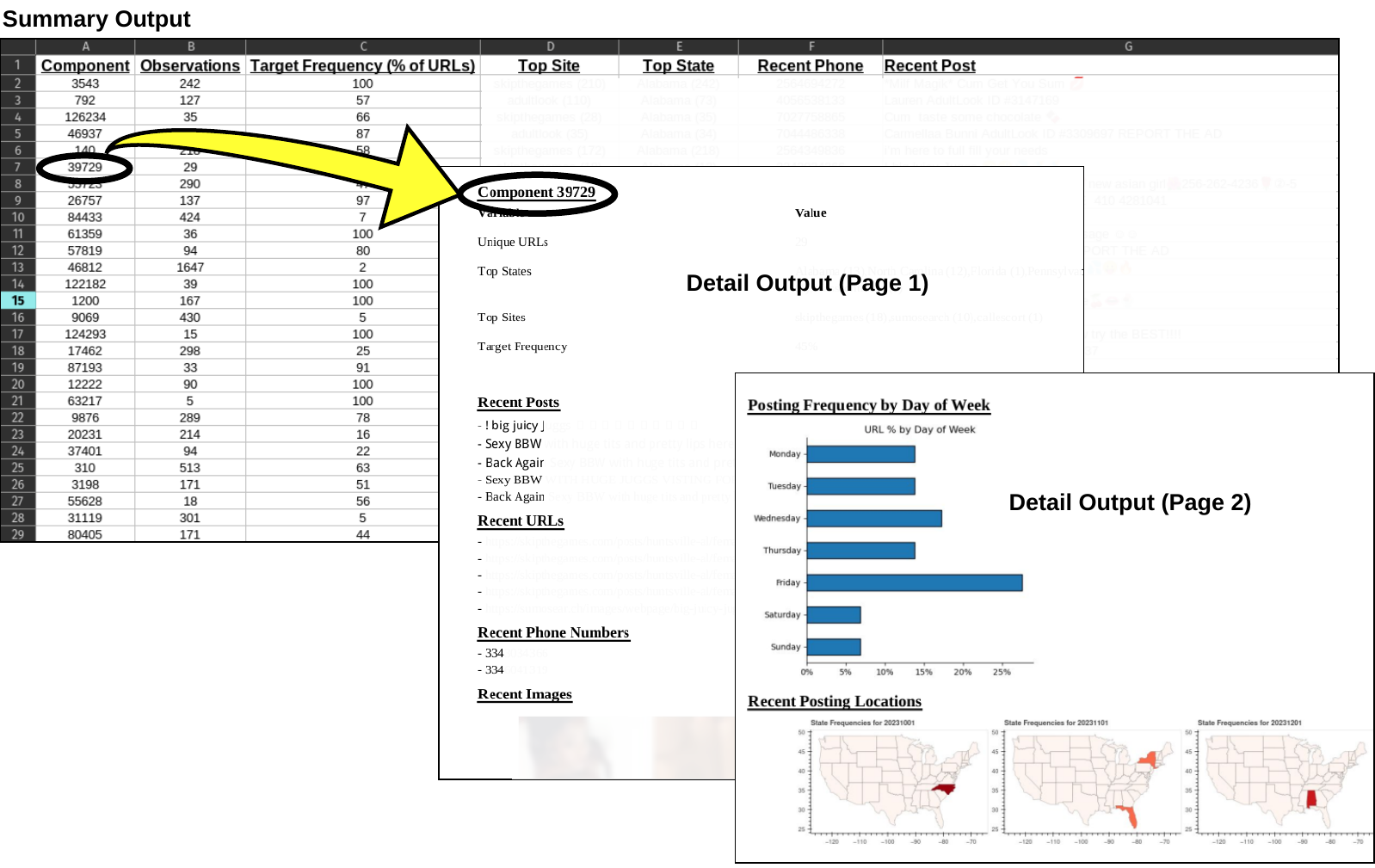}}
{Outputs Example\label{fig:output_example}}
{}
\end{figure}

The final phase of our ad-linking process generates outputs for our LEO and NPO partners engaged in counter-trafficking operations. We generate two types of outputs that are tailored to the region where our partners operate. The first, \textit{summary output}, provides a tabular listing of individuals or unique posting entities (corresponding to connected components in the graph representation) that have targeted the selected region in recent days. This output contains basic statistics such as the number of observed ads, the proportion of times the individual targeted the chosen region, and recently used contact information for each individual. This allows users to quickly browse the list and attempt to make contact with individuals without visiting the sites hosting ads, minimizing exposure to explicit content. This is an important consideration for our NPO partners who rely on volunteers for outreach efforts, including volunteers who are former victims of sex trafficking. The second, \textit{detail output}, provides additional details for each individual in the summary output, including recently posted images and post texts, recent URLs, and a choropleth plot showing recent movement across states by month. Our process allows specific details of the outputs to be specified on a per-partner basis, including the blur applied to images, the locations considered, and a minimum target ad frequency for the selected locations. Figure \ref{fig:output_example} provides an example of both outputs and shows how the summary and detail outputs are linked by the unique identifier associated with connected components of the graph representation. 

We do not discuss the specifics of how our partners operationalize these data products because they vary greatly depending on the type of organization (LEO vs. NPO) and the resources available. However, the key observation is that these data products distill data collected from large volumes of ads posted on one or more sites into interpretable information that can be used directly to initiate contact with a potential victim. Without these products, organization employees must try to establish links between ads by visiting sites and manually linking the information. Just the process of linking ads, excluding attempts to eliminate erroneous connections, is daunting. For example, if we observe 100 ads targeting a specific location, does that correspond to 100 individuals each posting a single ad, one individual posting 100 times, or some combination in between?

\subsection{Technologies}\label{subsec:technologies}
We implement the ad-linking process using the Python programming language. Table \ref{tab:libraries_by_task} specifies essential libraries, the tasks performed by each library, and any available citations. We use the following abbreviations for the tasks: DC - Data Collection, DCS - Data Cleaning and Standardization, LCS - Location Cleaning and Standardization, AG - Ad Grouping, and OG - Output Generation.

\begin{table}[htbp]
    \centering
    \begin{footnotesize}
    \begin{tabular}{lcccccp{8cm}}
         Library                & DC & DCS & LCS & AG & OG & Citation(s)  \\
         \hline
         \texttt{beautifulsoup} & X  &     &     &    &    \\
         \texttt{bokeh}         &    &     &     &    &  X \\
         \texttt{igraph}        &    &     &     &  X  &  & \cite{igraph}\\
         \texttt{matplotlib} &    &     &     &   & X & \cite{Hunter:2007} \\
         \texttt{numpy}   & X & X & X & X & X & \cite{harris2020array}\\
         \texttt{pandas} &  & X & X & X & X & \cite{reback2020pandas,mckinney-proc-scipy-2010} \\
         \texttt{pillow} &  X  &     &     &    &  X \\
         \texttt{polars} &  & X & X & X & X \\
         \texttt{python-docx} &    &     &     &    &  X \\
         \texttt{requests} &  X  &     &     &    &  \\
         \texttt{scikit-learn}  &    &     &     &  X  &  & \cite{scikit-learn}\\
         \texttt{scipy}  &    &     &     &  X  &  & \cite{scipy}\\
         \texttt{seaborn}  &    &     &     &    & X &  \cite{seaborn}\\
         \texttt{selenium} & X  &     &     &    &    \\
         \texttt{sentence-transformers} &    &     &     &  X  & & \cite{transformers}  \\
         \hline
    \end{tabular}
    \end{footnotesize}
    \caption{Python Libraries Used by Task}
    \label{tab:libraries_by_task}
\end{table}
\section{Giant Component Filtering}\label{sec:gc_edge_removal}
This section provides additional details regarding our procedure for detecting and removing potentially erroneous links in the graph representation of the giant component. As described in Section \ref{subsec:ad_grouping}, the procedure iterates over all edges in a projection of the giant component's graph representation, where the projection includes edges between posts that were initially connected by vertices corresponding to pHash or contact information values. During each iteration, a binary classification model assesses the probability that the same user created the posts associated with the connected vertices. Algorithm \ref{alg:gc_filtering} provides pseudocode for the filtering procedure.

\begin{algorithm}[htb]\label{alg:gc_filtering}
\caption{Giant Component Filtering}
\vskip6pt
\begin{small}
\begin{algorithmic}
\State - Define same user probability threshold, $\delta$.
\State - Extract connected components from graph representation of ASW data.
\State - Use sentence transformer (see Section \ref{sec:language_model}) and non-giant components to train a binary classifier for the same user classification task (see Section \ref{sec:binary_classification}).
\State - Define {\texttt{GC\_graph} as the giant component from the original graph representation}.
\State - Create post projection of {\texttt{GC\_graph} and store as \texttt{GC\_post\_graph}}. Let $E$ denote the set of edges in \texttt{GC\_post\_graph}
\State - Define $SU(x, y)$ as a procedure that computes the same user probability for a set of texts, $x$ and $y$, i.e., our binary classification model.
\State - Define an empty list, \texttt{edges\_to\_remove}.
\For{$e \in  E$}
    \State Let \texttt{source} and \texttt{target} represent the vertices connected by edge $e$ in \texttt{GC\_post\_graph}.
    \State Let $\tilde{s}$ represent the post text associated with the \texttt{source} vertex.
    \State $\tilde{t}$ represent the post text associated with the \texttt{target} vertex.
    \If{$SU(\tilde{s}, \tilde{t}) < \delta$}
        \State Add edge $e$ to \texttt{edges\_to\_remove}.
    \EndIf
\EndFor
\State - Remove edges in \texttt{edges\_to\_remove} from \texttt{GC\_post\_graph}.
\State - Extract connected components from \texttt{GC\_post\_graph}. Each connected component contains a set of vertices corresponding to posts that are \textit{reachable} to one another in \texttt{GC\_post\_graph}.
\State - Create a unique identifier for all extracted components and use these identifiers to label posts in the data.
\end{algorithmic}
\end{small}
\end{algorithm}

The remainder of this section focuses on the developed classification procedure, including how it is trained and deployed. We also present experimental results demonstrating the efficacy of the edge removal procedure on an open dataset that includes data from multiple ASWs. We begin with a description of the ASW dataset used throughout this section and a brief overview of the computational setup.

\subsection{Data}\label{sec:data}
We utilize the open dataset described in \cite{Freeman-2025-multisite} to demonstrate relevant aspects of our giant component filtering approach. The dataset provides a sample of more than 10 million ads collected from nine ASWs between May 1, 2022, and August 1, 2022. The authors obfuscate numerical data included in post texts, anonymize URLs and site names, and provide pHash values instead of images. These measures prevent the direct identification of individuals while preserving as much data structure as possible. Table \ref{tab:data_columns} lists the specific data fields we use and describes each field exactly as provided in the original paper. We refer readers seeking an in-depth discussion of the data collection methodology and structure to the original publication.

\begin{table}[htb]
    \centering
    \begin{tabular}{lp{13.5cm}}
    \hline 
    Data Field & Description \\ \hline
    \texttt{url} & An integer representing the URL associated with the ad. \\
    \texttt{site} & An integer representing the site the data was collected from. \\
    \texttt{post\_masked} & The raw heading text associated with the ad. \textbf{Note}: Numbers that appear in the text are obfuscated by replacing them with the character `*'. \\
    \texttt{post\_int} & An integer representing the raw heading text associated with the ad. \textbf{Note}: All occurrences of a specific text are assigned to the same integer value. \\
    \texttt{phone\_int} & An integer representing the phone number associated with the ad. Missing values are represented as \texttt{None}. \\
    \texttt{phash16} & The pHash for a single image associated with the ad, computed using the \texttt{imagehash} library. \\ \hline
    \end{tabular}
    \caption{Relevant Data Columns}
    \label{tab:data_columns}
\end{table}

Table \ref{tab:scientific_data_stats} summarizes the data for each of the nine sites included in the dataset. We also include an overall summary (``All'') corresponding to the case where all data is used. Even though the data was collected from May 1, 2022, to August 1, 2022, we note substantial differences in the number of URLs and unique post, pHash, and phone number values across the sites. Moreover, the summary statistics indicate substantial variation across sites concerning data duplication. For example, notice that although site 1 has less than 10\% of the number of URLs as site 9, site 1 has more unique post, pHash, and phone number values. Ultimately, the summary statistics highlight the degree of variation in data available from ASWs.

\begin{table}[htb]
    \centering
    \begin{tabular}{lrrrr}
        Site & URLs & Unique Posts & Unique pHashes & Unique Phones \\
        \hline
        Site 1 & 461,234 & 283,845 & 1,986,408 & 143,226 \\
        Site 2 & 257,252 & 105,400 & 625,396 & 257,250 \\
        Site 3 & 1,794,625 & 263,084 & 978,388 & 265,552 \\
        Site 4 & 121,199 & 13,909 & 148,812 & 6,383 \\
        Site 5 & 27,818 & 16,049 & 65,533 & 5,139 \\
        Site 6 & 1,592,738 & 215,664 & 1,371,934 & 154,126 \\
        Site 7 & 1,070,044 & 320,347 & 1,117,073 & 305,370 \\
        Site 8 & 169,114 & 146,022 & 2,000,931 & 3,688 \\
        Site 9 & 4,784,510 & 273,264 & 501,156 & 17,586 \\
        All & 10,278,534 & 1,036,868 & 6,219,151 & 497,104 \\
        \hline
        \end{tabular}
    \caption{Dataset Summary}
    \label{tab:scientific_data_stats}
\end{table}

\subsection{Computational Setup}\label{sec:computational_setup}
The experimental results that follow show that our ad-linking process is computationally efficient and can handle practical data volumes in a time that allows for daily updates. However, processing large volumes of ad data and running neural network inference on hundreds of thousands of posts requires high-performance computational infrastructure. We run the ad-linking pipeline on a Dell Precision 7865 workstation with a 64-core AMD Ryzen Threadripper CPU, 512 GB of RAM, and an NVIDIA RTX 6000 Ada GPU with 48 GB of memory. We use the \href{https://system76.com/pop/}{\textit{Pop!\_OS}} Linux distribution and the \href{https://pixi.sh/latest/}{\textit{Pixi}} package manager to build and manage reproducible Python environments with GPU support. All experimentation presented in this section was run on the described workstation.

\subsection{Language Modeling}\label{sec:language_model}
We use a binary classification model to filter edges in the giant component. Specifically, the binary classifier is trained to predict whether a pair of posts can be attributed to the same individual or posting entity. Constructing an effective classifier requires a high-quality means of assessing the similarity of ASW post texts. We leverage a sentence transformer model designed explicitly for ASW data that is proposed in \cite{freeman2025languagemodel} for our classification task. Specifically, we use the sentence transformer fine-tuned from the  \texttt{BERT-New30522-MLM} variant trained for 20 epochs.

\begin{figure}[htbp]
\FIGURE
    {
    \includegraphics[width=\textwidth]{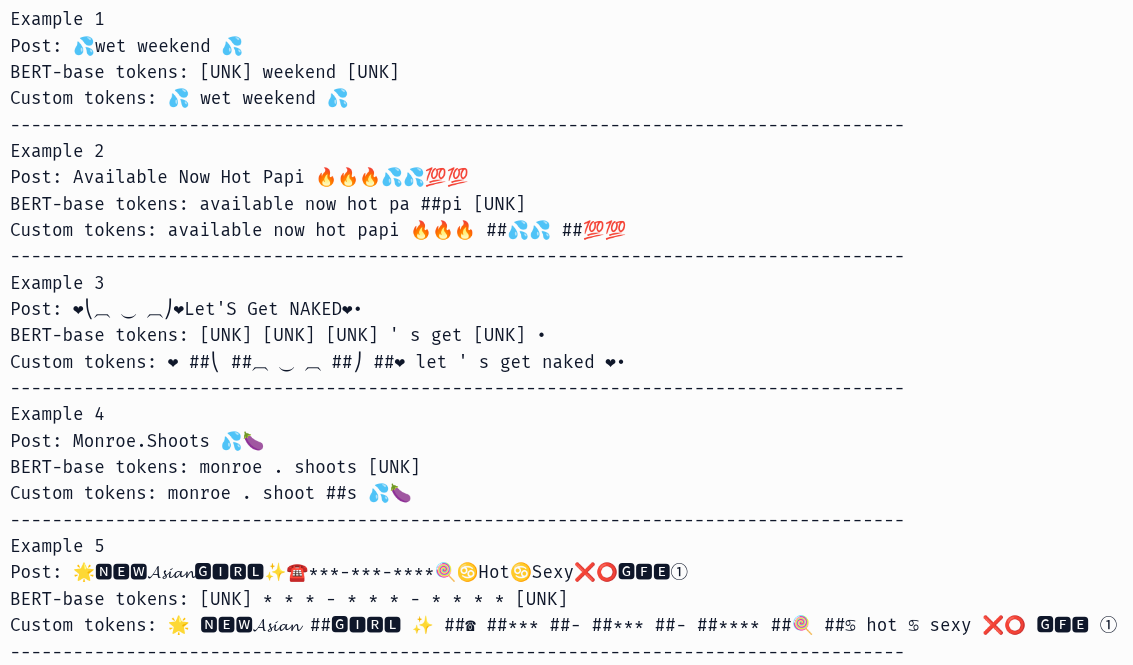}
    }   
{Tokenization Examples\label{fig:tokenization_examples}}
{}
\end{figure}

Figure \ref{fig:tokenization_examples} demonstrates the benefit of using the ASW-tailored models in terms of tokenization. The figure shows five example posts from the dataset described in Section \ref{sec:data} and how they would be tokenized by the \textit{base} variant of \texttt{BERT} (BERT-base) and the ASW-specific model from \cite{freeman2025languagemodel} (Custom). The key observations are that the BERT-base tokenization approach: ($i$) ignores emojis altogether, and $ii$) is more likely to be missing tokens in its vocabulary, which results in more occurrences of the \texttt{[UNK]} token.

\begin{figure}[htbp]
\FIGURE
{
    \includegraphics[width=\textwidth]{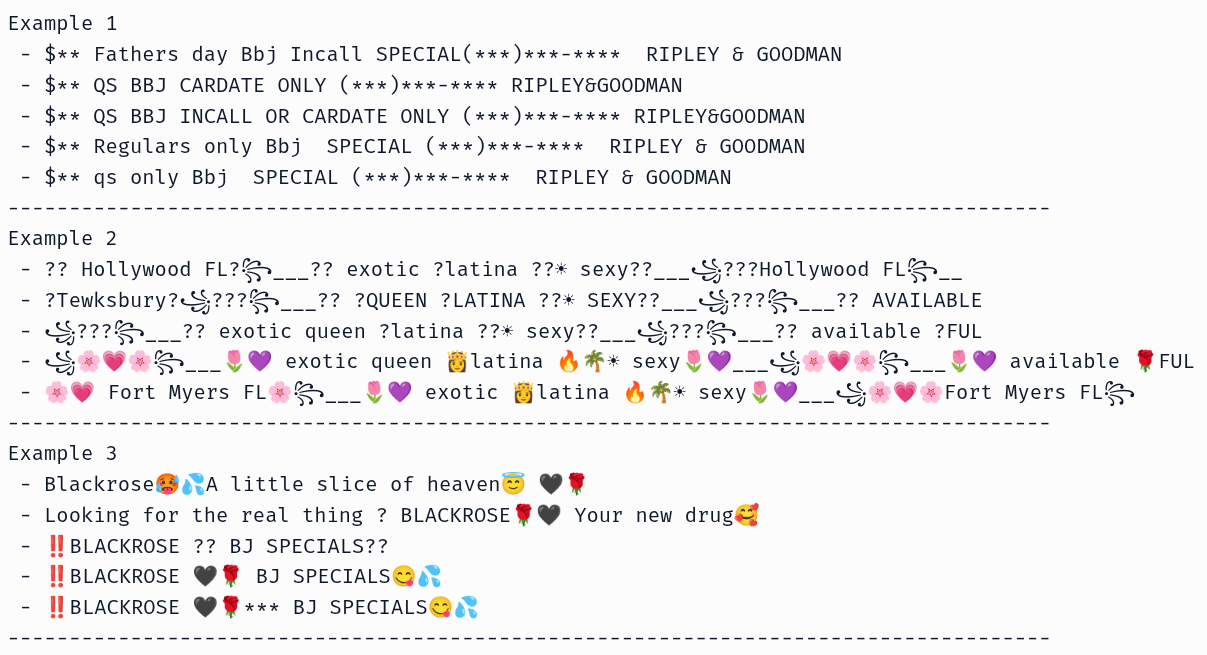}
}
{Embedding Cluster Examples\label{fig:cluster_examples}}
{}
\end{figure}

The previous figure demonstrates that the custom foundation models introduced in \cite{freeman2025languagemodel} are much better at retaining information from ASW post text values when tokenizing. We now demonstrate the sentence transformer model's ability to capture the semantic similarity of posts. Specifically, we use the trained sentence transformer model to generate embeddings for each unique post text value in our dataset. We then use the recently developed LocalMAP dimensionality reduction technique \citep{wang2024localmap} to reduce the 512-dimensional embeddings for each post text to a two-dimensional vector. We use the default values for LocalMAP when performing the dimensionality reduction. We then use HDBSCAN \citep{mcinnes2017hdbscan} to cluster the reduced-dimensional representations. Figure \ref{fig:cluster_examples} shows examples of posts associated with three resulting clusters, demonstrating the model's ability to generate embeddings that map similar texts nearby in the vector space.

\subsection{Binary Classification}\label{sec:binary_classification}
This section describes our approach to classifying whether pairs of post texts can be attributed to the same individual or posting entity. The classification task involves: ($i$) using the sentence transformer model described in the previous section to generate embeddings for pairs of connected posts in the projected giant component representation, ($ii$) computing the cosine similarity of the embeddings for linked posts to measure their proximity to the embedding space, and ($iii$) using a simple logistic regression model to learn the classification boundary that best distinguishes posts made by the same individual or entity.

\begin{figure}[htbp]
\FIGURE
{\includegraphics[width=\textwidth]{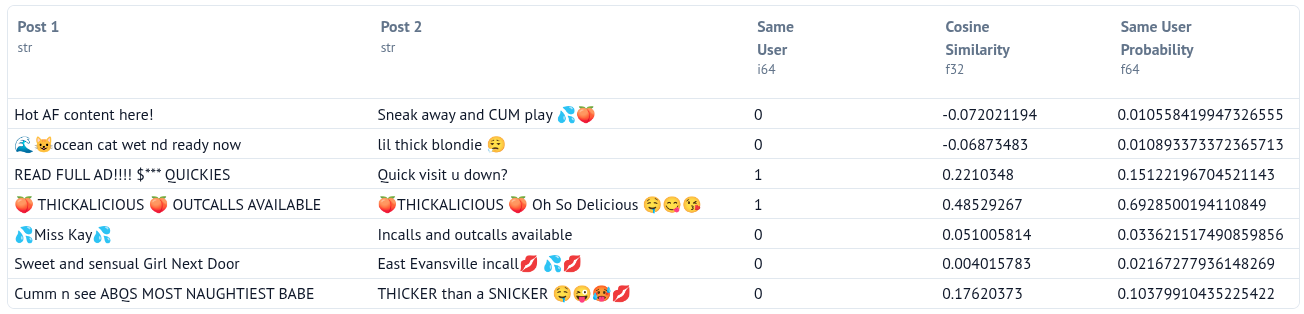}}
{Training Data Example\label{fig:training_df_example}}
{}
\end{figure}

The key challenge in training such a classifier is obtaining training data. Even though some sites may provide a profile identifier with ads, it is not common, and there is no way to guarantee that an individual or posting entity does not have multiple profiles. In short, ASWs generally do not provide a means to obtain ground truth labels for individuals posting ads. We overcome this challenge by assuming that non-giant components in the initial graph representations provide a proxy for assessing the expected similarity of posts for unique posting entities. Thus, we use the non-giant components from the original graph representation to construct the training dataset for the binary classification task. In particular, we: ($i$) extract the non-giant components each time the ad-linking process executes, ($ii$) use components with more than one linked post to generate examples of posts associated with the same individual or unique posting entity, i.e., positive class examples, and ($iii$) use post pairs selected from different components to create examples of posts associated with different posting entities, i.e., negative class examples. For all pairs in the training data, we generate text embeddings using the sentence transformer model and compute the cosine similarity between the embeddings. Since we know whether the post pairs in each row were selected from the same non-giant component, a binary label is added. We then train the logistic regression model to use the cosine similarity values to predict the label. In the giant component filtering procedure, we use the trained logistic regression model to generate the probability that a post pair can be attributed to the same posting entity, rather than simply predicting a class label. This allows us to control the tradeoff between false positives and false negatives. Figure \ref{fig:training_df_example} shows an example of the training dataset, along with computed cosine similarity and same user probability values.

\begin{figure}[htbp]
\FIGURE{
    \includegraphics[width=\textwidth]{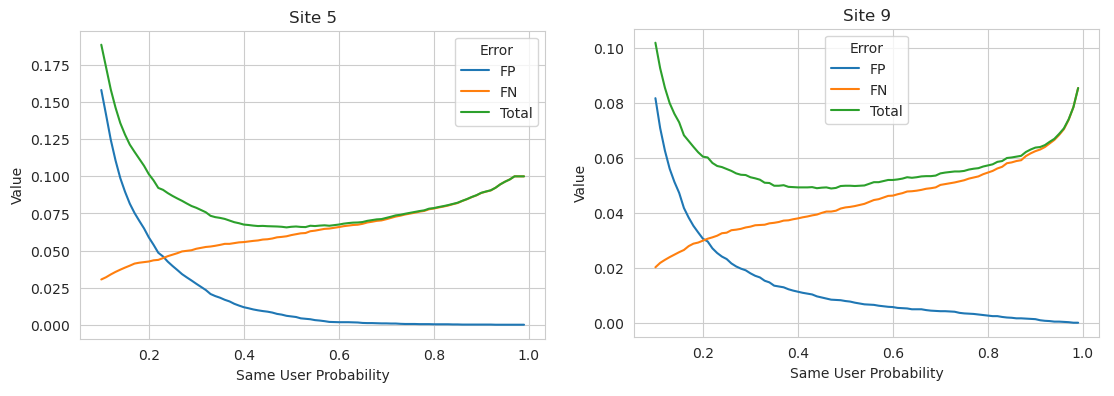}
}{Prediction Threshold Examples \label{fig:prediction_threshold_examples}}{}
\end{figure}

When generating the training dataset, we typically create 10,000 training examples, with 1,000 positive instances and 9,000 negative instances. Figure \ref{fig:prediction_threshold_examples} provides additional insight regarding how the selection of the same user probability threshold affects the rate of false positives and false negatives. The figure includes subplots that depict the tradeoffs between the false positive rate (FP), the false negative rate (FN), and the total error rate (Total) as the threshold for the same user probability varies from 0.0 to 1.0 along the horizontal axes. The left subplot shows the tradeoffs for site 5 (the smallest site by unique URLs), and the right subplot shows the tradeoffs for site 9 (i.e., the largest site by unique URLs).

Note in Figure \ref{fig:prediction_threshold_examples} that the errors are biased towards false negatives for probability thresholds greater than 0.25. In the context of our prediction task, a false negative (positive) corresponds to incorrectly leaving (removing) an edge in the giant component projection that should be removed (retained). Regardless, the same user probability threshold provides an easy-to-interpret mechanism that allows for tailoring potential errors made by the classification model based on user preferences. In practice, we use a threshold value of 0.7, which reduces false positives but increases the number of false negatives. However, the overall accuracy is comparable to that achieved with a threshold of 0.5, and the threshold can be easily adjusted based on partner preference.

\subsection{Impacts on Giant Component Size}\label{subsec:gc_size_impacts}
This section demonstrates the efficacy of our giant component filtering approach. We run the procedure for each site included in our dataset independently, as well as for all sites combined (``All''). We use a same user probability threshold of 0.7 when determining edges to remove in the giant component projection. Table \ref{tab:gc_filtering_summary} provides relevant statistics for this experiment, including:
\begin{hitemize}
    \item $|V|$ - the number of vertices in the original graph representation of the data,
    \item $|E|$ - the number of edges in the original graph representation of the data,
    \item $|C|$ - the number of components in the original graph representation of the data, including the giant component,
    \item GC Proportion - the size of the giant component as a proportion of the number of vertices ($|V|$),
    \item Filtered $|C|$ - the number of components after applying the giant component filtering procedure,
    \item Filtered GC Proportion - the size of the giant component as a proportion of the number of vertices ($|V|$) after applying the giant component filtering procedure,
    \item $|C|$ increase - the increase in the number of components due to the application of the giant component filtering procedure, expressed as a proportion of the number of components in the original graph representation ($|C|$).
\end{hitemize}

\begin{table}[htbp]
    \centering
    \begin{tabular}{lrrrcrcc}
         &  & & & & \multicolumn{1}{c}{Filtered} & \multicolumn{1}{c}{Filtered} & \\
        \multicolumn{1}{c}{Site} & \multicolumn{1}{c}{$|V|$} & \multicolumn{1}{c}{$|E|$} & \multicolumn{1}{c}{$|C|$} & \multicolumn{1}{c}{GC Proportion} & \multicolumn{1}{c}{$|C|$} & \multicolumn{1}{c}{GC Proportion} & \multicolumn{1}{c}{$|C|$ Increase} \\
        \hline 
        Site 1 & 2,413,478 & 3,740,639 & 69,990 & 0.23 & 93,370 & 0.02 & 0.33 \\
        Site 2 & 988,045 & 1,158,752 & 61,061 & 0.47 & 79,009 & 0.25 & 0.29 \\
        Site 3 & 1,507,023 & 2,329,791 & 56,715 & 0.50 & 100,682 & 0.21 & 0.78 \\
        Site 4 & 169,103 & 320,596 & 1,833 & 0.73 & 5,854 & 0.10 & 2.19 \\
        Site 5 & 86,720 & 194,188 & 3,738 & 0.31 & 6,294 & 0.02 & 0.68 \\
        Site 6 & 1,741,723 & 4,744,453 & 44,914 & 0.45 & 79,385 & 0.16 & 0.77 \\
        Site 7 & 1,742,789 & 2,708,611 & 62,769 & 0.53 & 119,987 & 0.22 & 0.91 \\
        Site 8 & 2,150,640 & 2,736,256 & 78,225 & 0.05 & 81,867 & 0.00 & 0.05 \\
        Site 9 & 792,005 & 1,007,027 & 82,841 & 0.08 & 95,751 & 0.02 & 0.16 \\
        All & 7,753,122 & 15,667,074 & 143,489 & 0.57 & 291,530 & 0.18 & 1.03 \\
        \hline
    \end{tabular}
    \caption{Giant Component Filtering Performance Summary}
    \label{tab:gc_filtering_summary}
\end{table}

The results show that our approach reduces the size of the giant component substantially for all cases. More importantly, the filtering approach results in the extraction of significant numbers of new components that can be used to generate actionable intelligence for potential victims of sexual exploitation. For the case where data from all sites is used to construct the initial graph representation, the giant component is reduced from a size corresponding to approximately 57\% of the data artifacts to 18\%, and the number of components is increased by more than 100\%. Without an effective means to decompose the giant component, all of the linked data would be disregarded because the high degree of noise prohibits actionable intelligence. Thus, the proposed approach is highly effective for increasing the amount of this important and hard-to-obtain data that can be acted upon.

\subsection{Benchmark Experiment}\label{sec:benchmarking}
We conclude this section with a benchmarking experiment that compares our giant component filtering to that described by \cite{freeman2022collaborating}. Specifically, we compare the two approaches in terms of reducing giant component size, elapsed time, and overall component increase.

\cite{freeman2022collaborating} use betweenness centrality (BC) to identify and remove nodes with unusually high connectivity. The BC value for a vertex $v$ captures the number of times $v$ lies on the shortest path between all other pairs of nodes. Thus, determining BC requires computing the shortest paths between all pairs of vertices in a graph. Noting this significant computational challenge, \cite{freeman2022collaborating} use a random sample of approximately 40\% of the vertices in a component to approximate the BC values for all vertices. They do not provide an exact BC threshold value that they use to identify vertices for deletion. Instead, they discuss how they represent BC values relative to the maximum observed value for different artifacts and examine changes in component structure at various relative cutoff thresholds to determine the appropriate settings.

Our benchmark experiment applies the giant component filtering approaches to our dataset. We apply the two approaches to each site individually and to all sites simultaneously (``All''). When implementing the approach from \cite{freeman2022collaborating}, which we refer to as the BC-based method, we: ($i$) construct the graph representation of the data, ($ii$) extract the giant component, ($iii$) approximate the BC values for all vertices in the giant component using a random sample of approximately 40\% of the giant component's vertices, ($iv$) express the BC values relative to the maximum value observed by artifact type, and ($v$) apply a percentile threshold to identify vertices for removal. For example, applying a threshold of 0.90 removes all vertices with a relative BC value that equals or exceeds the $90^{th}$ percentile for all artifact types. Noting that BC computation is expensive, we limit the time allowed to one day (86,400 seconds). Also, we use the \textit{nx-cugraph} library for Python to accelerate BC computation using the GPU available on the testing workstation (see Section \ref{sec:computational_setup}).

Table \ref{tab:benchmark_gc_reduction_time} presents the benchmark experiment results for giant component size reduction and elapsed time. We consider four different threshold values of 0.75, 0.9, 0.95, and 0.99 for the BC-based filtering procedure. The results associated with these thresholds are given in the columns BC - 0.75, BC - 0.9, BC - 0.95, and BC - 0.99, respectively. The reduction in giant component size is expressed as a percentage of the number of vertices in the original graph representation. The elapsed time is expressed in seconds. \textbf{Bold} font indicates the best performance for each site, with larger reductions in giant component size and shorter elapsed time values being preferred.

\begin{table}[htbp]
    \centering
    \begin{tabular}{|l|rrrrr|cc|}
    \hline
        \multicolumn{1}{|c}{} &
        \multicolumn{5}{|c|}{Giant Component Size Reduction} &
        \multicolumn{2}{c|}{Elapsed Time (Seconds)}\\
        \multicolumn{1}{|c}{Site} &
        \multicolumn{1}{|c}{Proposed} &
        \multicolumn{1}{c}{BC - 0.75} &
        \multicolumn{1}{c}{BC - 0.90} & 
        \multicolumn{1}{c}{BC - 0.95} &
        \multicolumn{1}{c|}{BC - 0.99} &
        \multicolumn{1}{c}{Proposed} & 
        \multicolumn{1}{c|}{BC} \\  
        \hline 
        Site 1 & \textbf{20.90\%} & 0.21\% & 0.20\% & 0.20\% & 0.20\% & \textbf{402} & 6,773 \\
        Site 2 & \textbf{22.90\%} & 16.77\% & 16.65\% & 16.65\% & 16.65\% & \textbf{275} & 5,802 \\
        Site 3 & 29.55\% & \textbf{50.39\%} & 50.36\% & 50.29\% & 49.62\% & \textbf{1,070} & 38,326 \\
        Site 4 & \textbf{62.55\%} & 10.51\% & 10.35\% & 10.35\% & 10.35\% & \textbf{82} & 3,257 \\
        Site 5 & \textbf{28.65\%} & 5.05\% & 5.03\% & 4.40\% & 4.40\% & 249 & \textbf{141} \\
        Site 6 & \textbf{28.83\%} & 0.08\% & 0.08\% & 0.08\% & 0.08\% & \textbf{1,039} & 7,863 \\
        Site 7 & \textbf{30.33\%} & 0.34\% & 0.16\% & 0.16\% & 0.16\% & \textbf{1,374} & 10,829 \\
        Site 8 & \textbf{4.85\%} & 0.04\% & 0.03\% & 0.03\% & 0.03\% & \textbf{74} & 803 \\
        Site 9 & \textbf{5.60\%} & 0.63\% & 0.63\% & 0.63\% & 0.63\% & \textbf{215} & 242 \\
        All & \textbf{39.16\%} & - & - & - & - & \textbf{9,300} & - \\
        \hline
    \end{tabular}
    \caption{Benchmark Experiment Results - Giant Component Reduction and Elapsed Time}
    \label{tab:benchmark_gc_reduction_time}
\end{table}

Table \ref{tab:benchmark_gc_reduction_time} shows that our giant component filtering procedure results in significantly larger size reductions when compared to that of \cite{freeman2022collaborating}. Moreover, our procedure tends to outperform in terms of elapsed time, with the BC-based procedure failing to complete within the one-day time limit for the ``All'' case. There are two exceptions: the BC-based method ($i$) yields a larger reduction in giant component size for Site 3, and $ii$) requires less computational time to filter the giant component for Site 5. We will see that the larger giant component reduction for Site 3 does not necessarily translate into a more \textit{effective} decomposition. Regarding the elapsed time result for Site 5, this site has the smallest graph representation (see Table \ref{tab:gc_filtering_summary}), suggesting that the computational overhead associated with computing post embeddings in our process can result in it being outperformed on small graphs. Table \ref{tab:benchmark_component_increase} compares the increase in components associated with the two approaches. \textbf{Bold} text to indicate the method that yields the largest increase in the number of components.

\begin{table}[htbp]
    \centering
    \begin{tabular}{|l|rrrrr|cc|}
    \hline
        \multicolumn{1}{|c}{} &
        \multicolumn{5}{|c|}{Component Increase}\\
        \multicolumn{1}{|c}{Site} &
        \multicolumn{1}{|c}{Proposed} &
        \multicolumn{1}{c}{BC - 0.75} &
        \multicolumn{1}{c}{BC - 0.9} & 
        \multicolumn{1}{c}{BC - 0.95} &
        \multicolumn{1}{c|}{BC - 0.99}\\  
        \hline 
        Site 1 & \textbf{33.41\%} & 2.95\% & 2.94\% & 2.94\% & 2.89\% \\
        Site 2 & \textbf{29.40\%} & 3.19\% & 1.70\% & 1.70\% & 1.70\% \\
        Site 3 & \textbf{77.52\%} & 0.08\% & 0.44\% & 1.35\% & 7.07\% \\
        Site 4 & 219.42\% & \textbf{602.45\%} & 602.40\% & 602.40\% & 602.40\% \\
        Site 5 & \textbf{68.41\%} & 1.69\% & 1.55\% & 1.47\% & 1.44\% \\
        Site 6 & \textbf{76.75\%} & 0.94\% & 0.94\% & 0.94\% & 0.94\% \\
        Site 7 & \textbf{91.16\%} & 5.63\% & 2.09\% & 2.09\% & 2.09\% \\
        Site 8 & \textbf{4.66\%} & 0.10\% & 0.05\% & 0.04\% & 0.01\% \\
        Site 9 & \textbf{15.59\%} & 0.67\% & 0.67\% & 0.67\% & 0.66\% \\
        All & \textbf{103.17\%} & - & - & - & - \\
        \hline
    \end{tabular}
    \caption{Benchmark Experiment Results - Component Increase}
    \label{tab:benchmark_component_increase}
\end{table}

Table \ref{tab:benchmark_component_increase} shows that our procedure typically results in substantial increases in the number of components. Recall that each component represents data associated with an individual or unique posting entity. Thus, the increase in components directly translates into a greater potential to identify potential victims of sexual exploitation. One exception is Site 4, where both methods yield substantial increases in components, but the BC-based approach increases the number of components by more than 600\%. Table \ref{tab:gc_filtering_summary} shows that the giant component in the graph representation for Site 4 includes approximately 75\% of all vertices, explaining the high potential for additional components. Moreover, the performance of the BC-based approach suggests it may be better for some site characteristics. 

Another interesting case is Site 3. While Table \ref{tab:benchmark_gc_reduction_time} showed that the BC-based method provides a larger reduction in giant component size for this site, our procedure is more effective at increasing the number of components. This observation highlights the potential for the BC-based approach to result in significant data loss. In other words, the substantial reduction in giant component size seems to be a result of deleting a significant proportion of vertices, which limits the increase in components. Our procedure does not delete data; instead, it removes data associations. Together, the results presented in Tables \ref{tab:benchmark_gc_reduction_time} and \ref{tab:benchmark_component_increase} show that our approach generally improves upon the performance of that described in \cite{freeman2022collaborating} in terms of giant component size reduction, computational time, and the degree of component increase. Moreover, it provides these benefits without data loss.
\section{Conclusion}\label{sec:conclusion}
We present an end-to-end process for collecting and linking sex ad data, identifying and removing erroneous links, and transforming this data into intelligence for counter-trafficking operations. Our methodology has been refined through ongoing collaborations with LEO and NPO partners, directly supporting operations that have identified and connected with more than 60 potential sex trafficking victims. Our ad-linking process addresses significant limitations in previous approaches, particularly that described by \cite{freeman2022collaborating}. A critical contribution is recognizing the emergence of a giant component in the graph representation of sex ad data, establishing an important connection between counter-trafficking efforts and network science. This recognition enables us to develop a tailored approach for detecting and removing potentially erroneous edges in the giant component. This procedure:
\begin{hitemize}
    \item is data-driven, leveraging recent advances in language models specifically tailored for ASW data,
    \item results in no data loss, preserving all collected information while removing potentially erroneous associations, and
    \item includes a novel approach to create training data based on observations from components not connected in the giant component, which allows our approach to evolve based on the available data.
\end{hitemize}

To aid reproducibility and validate our approach, we demonstrate its efficacy using an open dataset that contains a diverse sample of data collected from nine different ASWs. Our experimental results show substantial improvements over previous methods in terms of giant component size reduction, computational efficiency, and component increase, with our approach yielding an increase in the number of components exceeding 100\% when applied to the combined dataset. Moreover, we have taken care to explicitly describe our tools, techniques, and the reasoning behind the design choices embedded in the process.

Regarding limitations, first and foremost, we want to emphasize that not all ads posted on ASWs are associated with victims of sex trafficking, and our process does not attempt to predict whether or not an individual is being trafficked. Based on our fieldwork experience, we believe such a determination is impossible based solely on ASW data. Instead, our ultimate goal is to develop techniques that transform the large volumes of ASW data created daily into data products that enhance the efforts of subject matter experts in the field. 

Second, ASWs are only one area of the Internet that facilitates sex trafficking. Others include social media and sites on the Dark Web. However, given the well-documented history of sex trafficking being facilitated by ASWs and the general availability of these sites, we believe this work represents a significant advance in the fight against sexual exploitation.

Third, a key feature of our process is that it results in no data loss. However, we know that some proportion of ads posted on ASWs are scams \citep{freeman2022collaborating}. Instead of trying to detect them explicitly in our process, we allow them to remain. Table \ref{tab:gc_filtering_summary} shows that even though our procedure is very effective at extracting components from a giant component, it is often the case that some residual (but much smaller) giant component remains. It may be that the majority of scam ad data is contained in this residual giant component, or it may not. However, instead of making assumptions about how scam ads will manifest in the graph representation, we chose to allow them to persist. A better understanding of this manifestation is a fruitful area for future research. Note that when working with partners in the field, we apply filters to limit the components that are used to construct data products. One filter we commonly apply only includes components that have posted ads targeting the partner's location at a specified targeting rate or higher. Our experience suggests that scammers tend to target many locations simultaneously with their ads, which makes the targeting rate for any specific location rather low. Thus, our experience suggests the risk of scam ads making it to data products is low.

In addition to a better understanding of how scam ads manifest in the graph representation of ad data, two additional research directions come to mind. The first is a better understanding of the online commercial sex ecosystem and tools to quickly identify new sites and shifts in volume. As noted in the introduction, this ecosystem is highly fluid and subject to significant disruptions in response to policy changes. During our years working in this area, we have observed dramatic shifts in site usage following changes such as the implementation of more stringent account creation guidelines or the introduction of posting fees. We have also identified cases where posting entities in certain regions use specific sites or boards that are not widely used in other areas. Improving discovery mechanisms for these emerging patterns remains a challenge. Second, while we use images for linking advertisements, we do not currently use them to detect erroneous links. Although we have experimented with tools to utilize data embedded in images (such as facial recognition), many of these tools perform poorly with the low-resolution images prevalent on ASWs. Given the volume of advertisements hosted by these sites, the majority of images are of low resolution, necessitating the development of specialized tools and techniques that can accommodate this reality.

In conclusion, our ad-linking process represents a significant methodological advancement in the fight against sex trafficking, providing LEO and NPO partners with actionable intelligence derived from the vast amount of data available on ASWs. By connecting network science principles with counter-trafficking efforts and developing specialized tools for this domain, we have established a process that supports the efficient and effective identification of potential victims and provides a foundation for future research in this critical area.

% Appendix here
% Options are (1) APPENDIX (with or without general title) or
%             (2) APPENDICES (if it has more than one unrelated sections)
% Outcomment the appropriate case if necessary
%
% \begin{APPENDIX}{<Title of the Appendix>}
% \end{APPENDIX}
%
%   or
%
% \begin{APPENDICES}
% \section{<Title of Section A>}
% \section{<Title of Section B>}
% etc
% \end{APPENDICES}

% Acknowledgments here
\ACKNOWLEDGMENT{This work is partially supported by the National Science Foundation.}

% References here (outcomment the appropriate case)

% \bibliographystyle{informs2014} % outcomment this and next line in Case 1
% \bibliography{ad_linking} % if more than one, comma separated

%%%%%%%%%%%%%%%%%
\end{document}